\documentclass[10pt,aps,prl,twocolumn,superscriptaddress,showpacs,floatfix]{revtex4-1}

\usepackage{graphicx}
\usepackage{amssymb}
\usepackage{amsmath}
\usepackage{xspace}
\usepackage{color}
\usepackage{braket}

\begin{document}
\title{Order, criticality and excitations in the 
extended Falicov-Kimball model}

\author{S. Ejima}
\affiliation{Institut f\"ur Physik, Ernst-Moritz-Arndt-Universit\"at
Greifswald, 17489 Greifswald, Germany}

\author{T. Kaneko}
\author{Y. Ohta}
\affiliation{Department of Physics, Chiba University, Chiba 263-8522, Japan}

\author{H. Fehske}
\affiliation{Institut f\"ur Physik, Ernst-Moritz-Arndt-Universit\"at
Greifswald, 17489 Greifswald, Germany}

\date{\today}

\begin{abstract}
Using exact numerical techniques we investigate the nature of excitonic 
(electron-hole) bound states and the development of exciton coherence 
in the one-dimensional half-filled extended Falicov-Kimball model. 
The ground-state phase diagram of the model exhibits, besides band insulator 
and staggered orbital ordered phases, an excitonic insulator (EI) 
with power-law correlations. 
The criticality of the EI state shows up in the von Neumann entropy.
The anomalous spectral function and condensation amplitude 
provide the binding energy and coherence length of the electron-hole pairs which, on their part,  
point towards a Coulomb interaction driven crossover from BCS-like electron-hole pairing fluctuations 
to tightly bound excitons. We show that while a mass imbalance between electrons and
holes does not affect the location of the BCS-BEC crossover regime it favors 
staggered orbital ordering to the disadvantage of the EI. 
Within the BEC regime the quasiparticle dispersion develops  a flat valence-band top in accord with the experimental finding for 
Ta$_2$NiSe$_5$.
\end{abstract}


\maketitle

The formation and condensation of excitonic bound states of electrons and holes 
in semimetallic or semiconducting systems  possessing a small band overlap or band gap 
is still---half a century after its theoretical prediction~\cite{Mo61}---a topical issue 
in condensed matter physics~\cite{NW90,CMCBDGBA07,MBSM08}.
If the binding energy  of the excitons exceeds the overlap/gap they may spontaneously 
condensate at low temperatures and drive the system into an excitonic insulator (EI) 
state.  It has been pointed out that the semimetal-EI transition can be discussed in 
close analogy to the BCS superconductivity whereas the semiconductor-EI transition is 
described in terms of a Bose-Einstein condensation (BEC) of preformed excitons~\cite{BF06}.
Quite recently, as a candidate for the EI state, quasi one-dimensional (1D) 
Ta$_2$NiSe$_5$ has raised and attracted much experimental attention~\cite{WSTMANTKNT09}.
Most notably, by angle-resolved photoemission spectroscopy (ARPES), an
extremely flat valence-band top at 40 K
was observed and taken as a strong signature for the EI state to be formed  
out of `condensed' bound Ni 3$d$ -- Se 4$p$ holes and Ta 5$d$ electrons.

The detection of the EI state in Ta$_2$NiSe$_5$  has spurred multifaceted research activities
with regard to the formation and possible condensation of excitons
in 1D systems~\cite{KTKO13}. 
The minimal theoretical model in this respect is of the Falicov-Kimball type. 
While the original Falicov-Kimball model (FKM) describes  
localized $f$ electrons interacting via a local Coulomb repulsion ($U$)  with itinerant $c$ electrons ($t_c$) 
if residing at the same Wannier site~\cite{FK69}, an extended version takes into account 
also the direct nearest-neighbor  $f$-electron hopping ($t_f$)~\cite{Ba02b}:   
\begin{eqnarray}
 {\cal H}&=& -t_c\sum_{\langle i, j \rangle}
               c_{i}^{\dagger}c_{j}^{\phantom{\dagger}}
	     -t_f\sum_{\langle i, j \rangle}
               f_{i}^{\dagger}f_{j}^{\phantom{\dagger}}
             +U\sum_i 
                c_{i}^{\dagger}c_{i}^{\phantom{\dagger}}
                f_{i}^{\dagger}f_{i}^{\phantom{\dagger}}
\nonumber\\
	     && +\frac{D}{2}\sum_i 
                 \left(
                  c_{i}^{\dagger}c_{i}^{\phantom{\dagger}}
                 -f_{i}^{\dagger}f_{i}^{\phantom{\dagger}}
                 \right)\,.
\label{hamilEFKM}
\end{eqnarray}
Here $\alpha_{i}^{\dagger}$ ($\alpha_{i}^{\phantom{\dagger}}$) 
denotes the creation (annihilation) operator of a spinless fermion 
in the $\alpha=\{c,\ f\}$  orbital at site $i$,  and $D$
is the level splitting between different $\alpha$-orbitals. 
In regard to the modeling of  Ta$_2$NiSe$_5$, the half-filled band case is of particular importance, and it has been shown
theoretically that a direct $f$-$c$ hopping (hybridization) is prohibited by symmetry reasons, at least between the valence band top and conduction 
band bottom~\cite{KTKO13}.

For the original FKM rigorous results were obtained only 
in infinite spatial dimensions by dynamical mean-field theory (DMFT),
see, e.g., reviews in Refs.~\cite{FZ03,Ke94}.
The extended FKM (EFKM) [Eq.~\eqref{hamilEFKM}] has been studied extensively 
in the context of EI formation for D~$>1$, using 
DMFT~\cite{TKVL11},  random phase approximation~\cite{ZIBF12}, slave-boson~\cite{Br08}, projective renormalization~\cite{PBF10} 
and variational cluster~\cite{SEO11} techniques, or purely numerical 
diagonalization procedures~\cite{KEFO13}. 
At the same time, the problem of electronic ferroelectricity, which is equivalent 
to the appearance of the EI in some theoretical models, has also attracted 
much attention~\cite{POS96a,YMST11}.   This phenomenon  
was confirmed for the 2D EFKM by constrained path Monte Carlo simulations~\cite{BGBL04}.
In 1D, however,  true ferroelectric long-range order  (the equivalent of a nonvanishing $\langle c^\dagger f \rangle$ expectation value
in the limit of vanishing $c$-$f$ band hybridization) is not possible. This was demonstrated for the 1D FKM~\cite{Fa99}. 
For the 1D EFKM power-law critical (excitonic) correlations were  observed instead~\cite{BGBL04}.  Mean-field based 
approaches~\cite{SC08}  are unable to capture  the EI state in 1D (despite their success for D$>$1), 
mainly due to the lack of an order parameter  associated with the breaking of the $U(1)$ symmetry.
On this note a thorough investigation of the ground-state and spectral properties of the 1D EFKM is still missing.  
 
In  this paper we present a comprehensive numerical analysis of the 1D EFKM at half filling.
At first we determine the ground-state phase diagram from large-scale density matrix renormalization group (DMRG)~\cite{Wh92} 
calculations and identify--depending on  the orbital level splitting---staggered orbital ordered (SOO) and band insulator (BI) phases as well as  
an intervening critical EI state.  Then, within the EI, we detect a crossover between BCS- and Bose-Einstein-type condensates  
monitoring the exciton-exciton correlation and exciton momentum distribution functions.  Note that in our 1D setting we use the term `condensate' 
to indicate a critical phase with power-law correlation decay. Finally,
combining DMRG, Lanczos exact diagonalization (ED) and Green functions 
techniques~\cite{OSEM94}, we study the anomalous spectral function and extract the correlation 
length and binding energy of the electron-hole pairs.  This allows us to comment on the nature of the excitonic 
bound states preceding the condensation process and to discuss the effect of a mass imbalance between ($c$-) electrons and ($f$-) holes. 

Examining the (large-$U$) strong-coupling regime gives a first hint 
of which phases might be realized in the 1D EFKM at zero temperature.  
To leading order the EFKM can be mapped onto the exactly solvable 
spin-1/2 XXZ-Heisenberg model in a magnetic field $h=D$ 
aligned in the $z$-direction~\cite{FDL95}, 
$
{\cal H}_{\rm XXZ}
 =J\sum_{j}
    \left\{\Delta S_j^z S_{j+1}^z
     +(1/2)(S_j^+S_{j+1}^- + S_j^- S_{j+1}^+)
    \right\} -h\sum_{j} S_j^{z}
$
with $J=4|t_f|t_c/U$ and $\Delta=(t_f^2+t_c^2)/(2|t_f|t_c)$. The XXZ model exhibits
three phases: the gapped antiferromagnetic (AF) phase, the critical gapless XY phase with central charge $c=1$,
and the ferromagnetic (FM) phase, where both transition lines, those between  AF and XY phases ($h_{{\rm c}_1}/J$)
and those between  XY and FM phases  ($h_{{\rm c}_2}/J$),   
follow from the Bethe ansatz~\cite{CG66}. 
Correspondingly,  increasing the magnitude of the $f$-$c$ level splitting $D$ in the EFKM, we expect to find the following 
sequence of phases: (i) the SOO phase that matches the Ising-like AF phase in the XXZ model, 
(ii) an intermediate critical EI phase with finite excitonic binding energy, and (iii) a BI state, 
which is characterized by a filled (empty) $f$ ($c$) band and related to the FM phase of the XXZ model.
The phase boundary separating the EI and BI states is exactly known to be~\cite{KS96b}
\begin{eqnarray}
 D_{{\rm c}_2}=\sqrt{4(|t_f|+|t_c|)^2+U^2}-U\,.
\label{Dc2-exact}
\end{eqnarray}

\begin{figure}[tbp]
 \begin{center}
 \includegraphics[clip,width=\columnwidth]{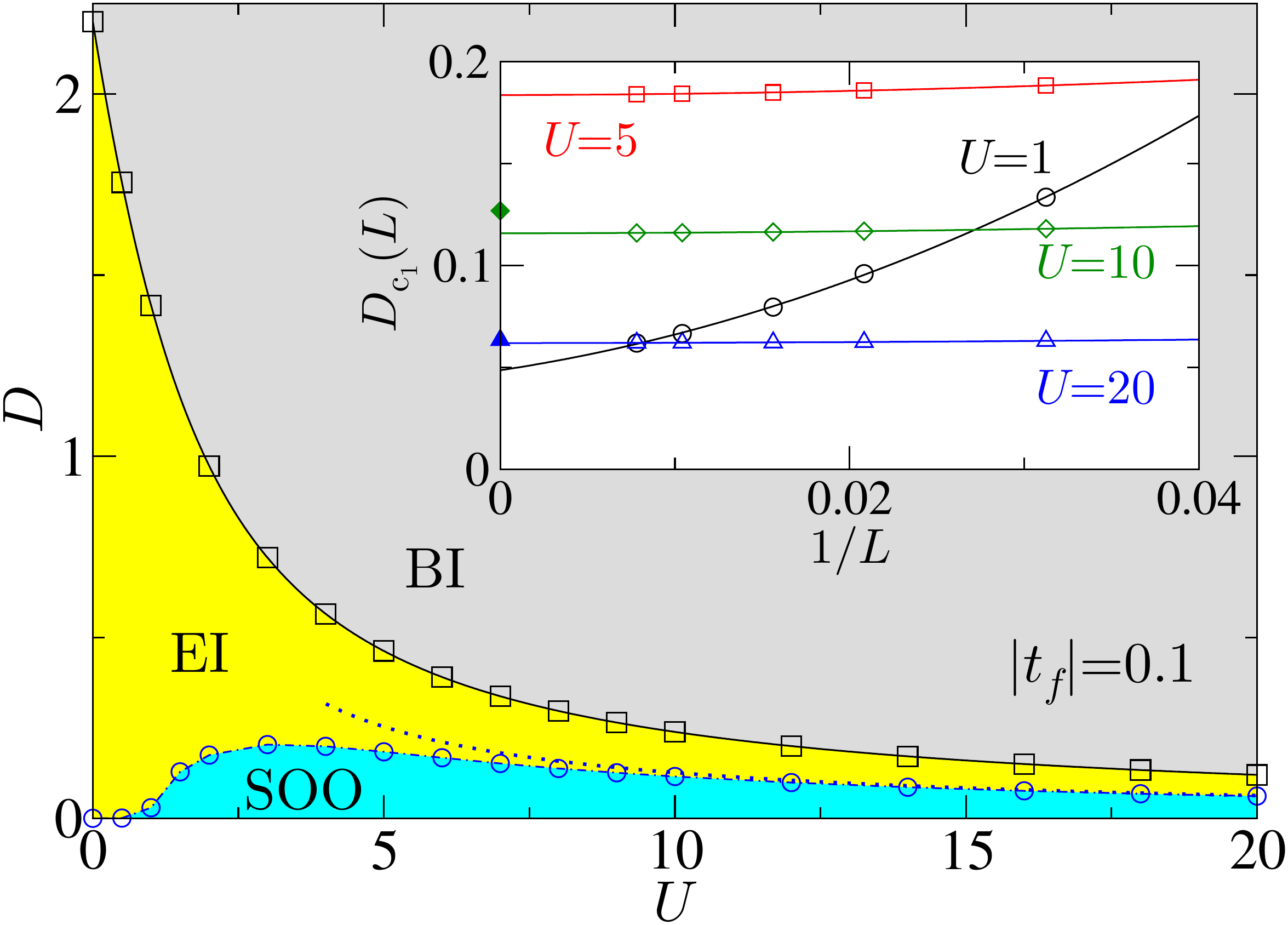}
 \includegraphics[clip,width=\columnwidth]{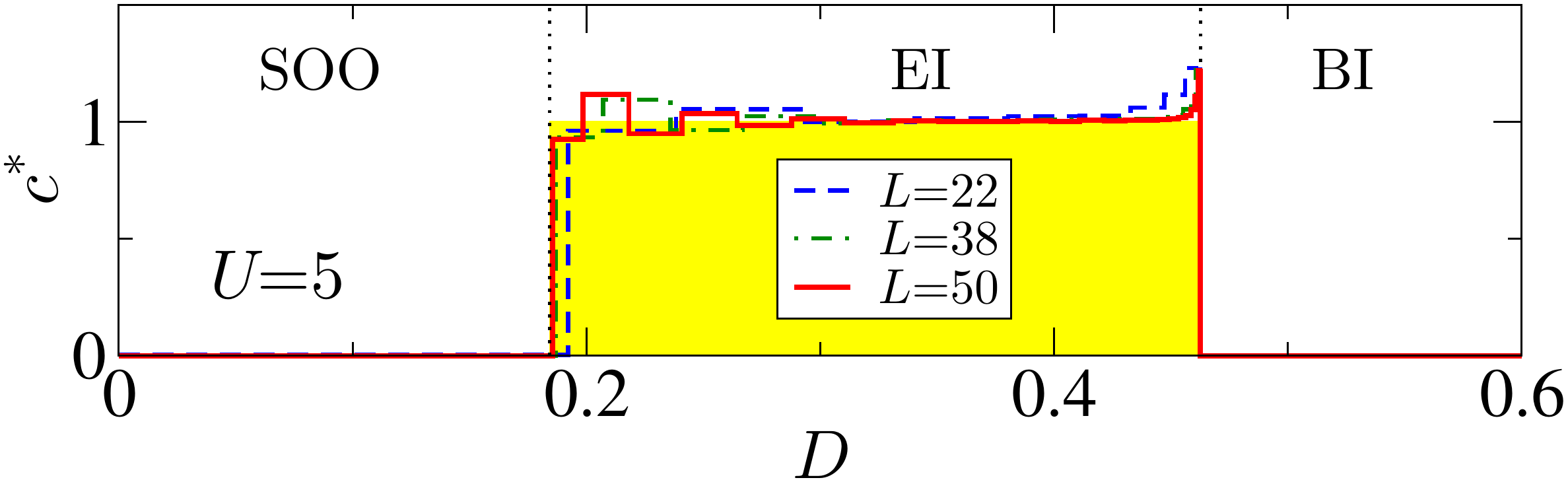}
 \end{center}
 \caption{(Color online) 
 Upper panel: Ground-state phase diagram of the half-filled 1D EFKM with $|t_f|=0.1$. Here and in what follows 
we take $t_c$ as the unit of energy. 
  Squares (circles)  denote the EI-BI (EI-SOO) 
 transition points $D_{{\rm c}_2}$ ($D_{{\rm c}_1}$) obtained by  DMRG method with up to $L=128$ sites and OBC.
 The solid line gives the analytical solution~\eqref{Dc2-exact}  
 for the EI-BI boundary;  the dotted line shows  the strong-coupling result for the EI-SOO 
 boundary.  The finite-size scaling of $D_{{\rm c}_1}(L)$ 
 is illustrated by the inset (open symbols), here the corresponding strong-coupling data
 are given by filled symbols.
 Lower panel: Central charge obtained at $U=5$ for various $L$ and PBC. Criticality, $c^\ast\sim1$,
 is observed for the EI. 
 }
\label{PD-tf01}
\end{figure}

The complete phase diagram of the 1D EFKM is presented in Fig.~\ref{PD-tf01}. 
Symbols denote the DMRG BI-EI and EI-SOO transition points, which can 
be obtained from the energy differences 
\begin{eqnarray}
 D_{{\rm c}_2}(L)=E_0(L,0)-E_0(L-1,1)=-E_0(L-1,1) 
 \label{Dc2}
\end{eqnarray}
and 
\begin{eqnarray}
 D_{{\rm c}_1}(L)=E_0(L/2+1,L/2-1)-E_0(L/2,L/2)\,,
 \label{Dc1}
\end{eqnarray}
respectively, in the course of a finite-size scaling analysis (see the inset). 
Here $E_0(N_f,N_c)$ denotes the ground-state energy
for a system with $N_f$ $f$- and $N_c$ $c$-electrons at $D=0$.
 Note that Eq.~\eqref{Dc2} 
holds for both, open and periodic boundary conditions (OBC/PBC), whereas
Eq.~\eqref{Dc1} has to be evaluated with PBC (if here OBC were used, an extra factor 2
results: $D_{{\rm c}_1}^{\rm OBC}=2D_{{\rm c}_1}$). For the DMRG runs performed in this work we keep at least  $m=3200$ density-matrix eigenstates which ensures 
a discarded weight smaller than $1\times 10^{-6}$.
The $D_{{\rm c}_2}(L\to\infty)$ values demonstrate the accuracy of our DMRG
calculations. Exact results for $D_{{\rm c}_1}(L\to\infty)$ can only be obtained numerically,
where a comparison with the dotted line reveals the limits of the strong-coupling approach~\cite{FDL95}; see Fig.~\ref{PD-tf01}. 
The criticality of the EI phase---corresponding to the critical XY phase 
in the XXZ model with  central charge $c=1$---can be confirmed by 
the von Neumann entanglement entropy $S_L(\ell)=-{\rm Tr} _{\ell}(\rho_\ell\ln\rho_\ell)$ 
(with reduced density matrix $\rho_\ell={\rm Tr}_{L-\ell}(\rho)$). 
Numerically, the central charge is best estimated from 
the entropy difference~\cite{CC04,Ni11}:
 \begin{eqnarray}
c^\ast(L)\equiv 3[S_L(L/2-1)-S_L(L/2)]/\ln\left[\cos(\pi/L)\right]\,. 
\end{eqnarray}
Our results for $c^\ast$, displayed in the lower
panel of Fig.~\ref{PD-tf01} for $|t_f|=0.1$ at $U=5$, give clear evidence that
$c^\ast\to  1$ in the EI, whereas we find $c^\ast=0$ in the BI and SOO phases. 
Regrettably, $c^\ast(L)$ is strongly system-size dependent near the EI-SOO transition.

\begin{figure}[tbp]
 \begin{center}
 \includegraphics[clip,width=\columnwidth]{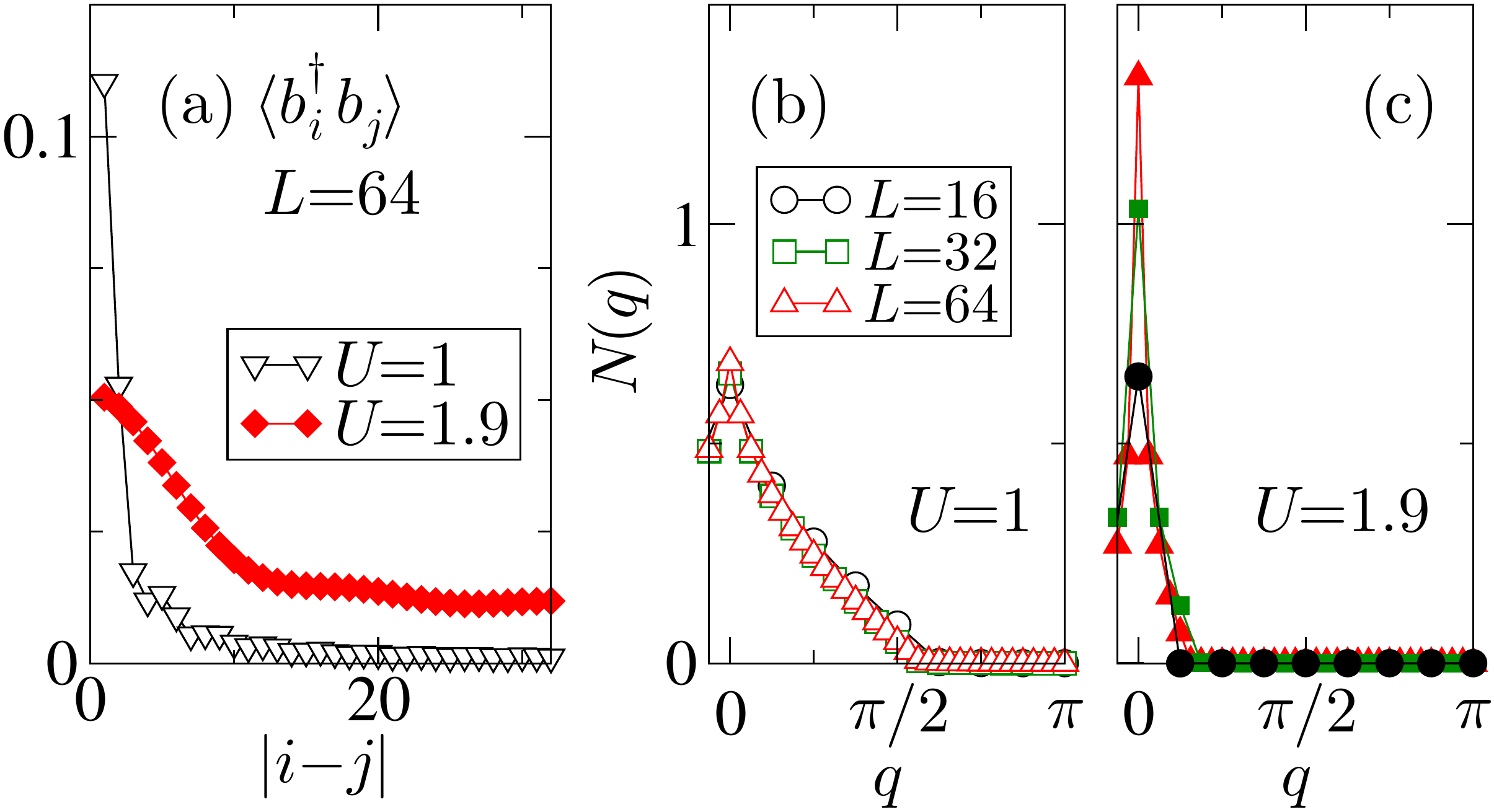}
 \end{center}
 \caption{(Color online) Exciton-exciton correlation function 
 $\langle b_{i}^{\dagger} b_{j}^{}\rangle$ (a)
 and excitonic momentum distribution function $N(q)$
 at $U=1$ (b) and $1.9$ (c) for $t_f=-0.1$, $D=1$.
 Data are obtained by the DMRG for 1D $L$-site lattices with PBC.
 } 
 \label{Nq}
\end{figure}

Let us now discuss the nature of the EI state in more detail. 
For simplicity  we consider the case  $t_f t_c<0$, where two Fermi points ($\pm k_{\rm F}$) exist 
for $U=0$ provided $D$ is sufficiently small (otherwise a direct band gap emerges).   
As a signature of an excitonic  Bose-Einstein condensate in 1D
one expects (i) a power-law decay of the correlations
$\langle b_{i}^\dagger b_{j}^{\phantom{\dagger}}\rangle$
with $b_{i}^\dagger=c_{i}^\dagger f_{i}^{\phantom{\dagger}}$
and (ii) a divergence of the excitonic momentum distribution
$N(q)=\langle b_{q}^\dagger b_{q}^{\phantom{\dagger}}\rangle$
with $b_{q}^\dagger=(1/\sqrt{L})\sum_k c_{k+q}^\dagger f_{k}^{\phantom{\dagger}}$
for the state with the lowest possible energy (in the direct gap case at $q=0$)
due to the absence of true long-range order. 
Figure~\ref{Nq} supports these expectations: Whereas in the weak-coupling
BCS regime ($U=1$), $\langle b_{i}^\dagger b_{j}^{\phantom{\dagger}}\rangle$
decays almost exponentially  and $N(q)$ shows only a marginal system-size dependence
(for all momenta), in the strong-coupling BEC regime close to 
the EI-BI transition ($U=1.9$), 
$\langle b_{i}^\dagger b_{j}^{\phantom{\dagger}}\rangle$ exhibits
a rather slow algebraic decay of the excitonic correlations
and $N(q=0)$ becomes  divergent as $L\to \infty$. 

We note that  the $\langle c^\dagger f\rangle$-expectation value is
always zero for a 1D system 
in the absence of an explicit $f$-$c$-band hybridization.
To examine the BCS-BEC crossover we adopt a technique introduced for detecting the particle 
fluctuations of Cooper pairs in 2D systems~\cite{OSEM94}.  
That is, we consider the off-diagonal anomalous exciton Green function
\begin{eqnarray}
 G_{cf}(k,\omega)
  =\Braket{
    \psi_1\left|c_k^\dagger\frac{1}{\omega+{\rm i}\eta-{\cal H}+E_0}f_k
   \right|\psi_0},
\end{eqnarray}
where $|\psi_0\rangle$ is the ground state $|N_f,N_c\rangle$
with fixed numbers of $f$- and $c$-electrons, 
$|\psi_1\rangle$ is the excited state $|N_f-1,N_c+1\rangle$,
$E_0$ is the averaged energy 
of $|\psi_0\rangle$ and $|\psi_1\rangle$, and $\eta$ is a broadening, 
and determine the corresponding spectral function  
$F(k,\omega)=(-1/\pi)\Im G_{cf}(k,\omega)$ that  gives the condensation amplitude 
$F(k)=\langle\psi_1|c^\dagger_{k}f^{\phantom{\dagger}}_k|\psi_0\rangle$.
 $F(k)$ can be directly computed by the ground-state DMRG method 
 taking into account  an extra target state $|\psi_1\rangle$. 
From $F(k)$ the coherence length characterizing  the excitonic condensate follows as
\begin{eqnarray}
 \xi^2=\sum_k |\nabla_k F(k)|^2\Big/\sum_k|F(k)|^2\,.
\end{eqnarray}
The binding energy of the excitons, $E_{\rm B}$, can be also determined from diverse ground-state energies~\cite{KEFO13}:
\begin{eqnarray}
 E_{\rm B}&=&E_0(N_f-1,N_c+1)+E_0(N_f,N_c)
  \nonumber \\
  && -E_0(N_f-1,N_c)-E_0(N_f,N_c+1)\,.
\label{E_B-static}
\end{eqnarray}

\begin{figure}[tbp]
 \begin{center}
 \includegraphics[clip,width=\columnwidth]{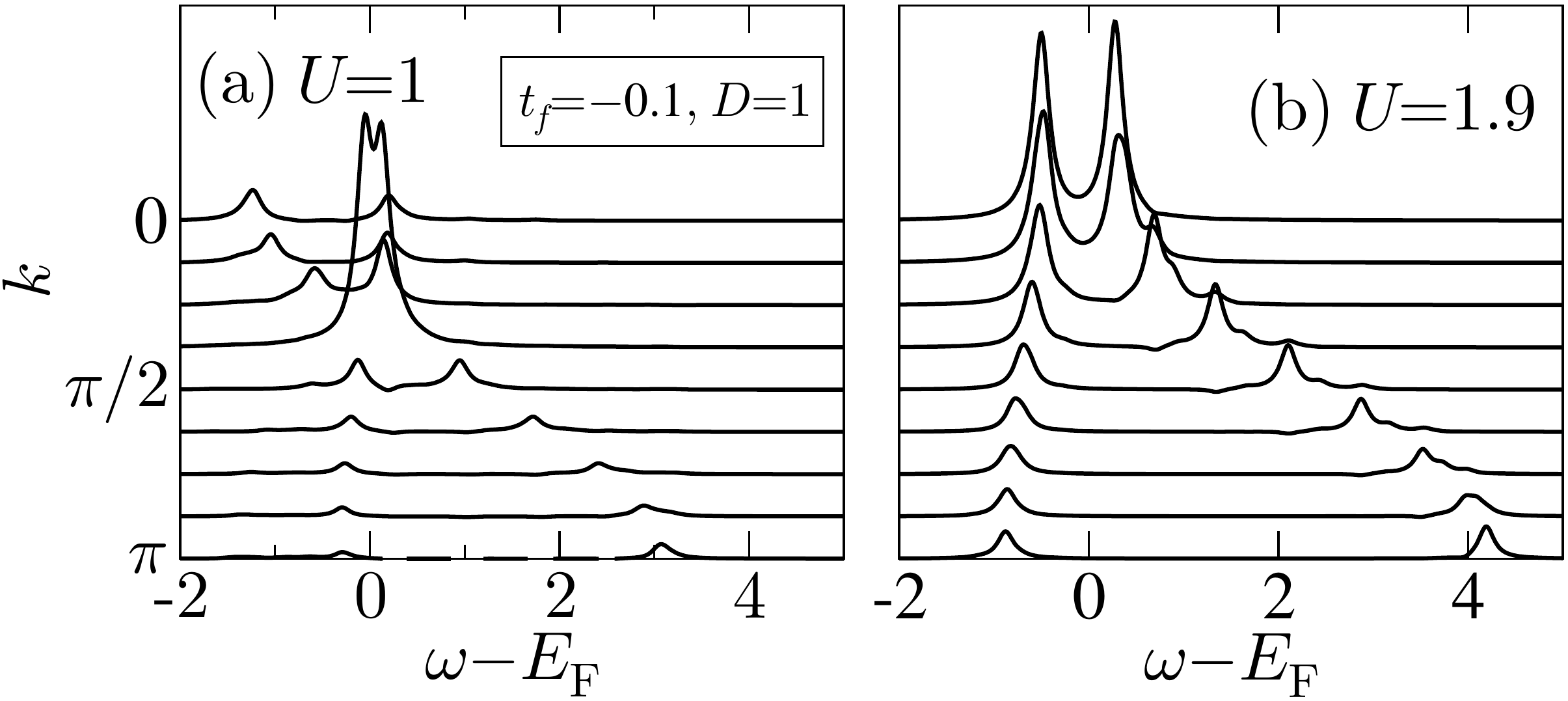}
 \includegraphics[clip,width=\columnwidth]{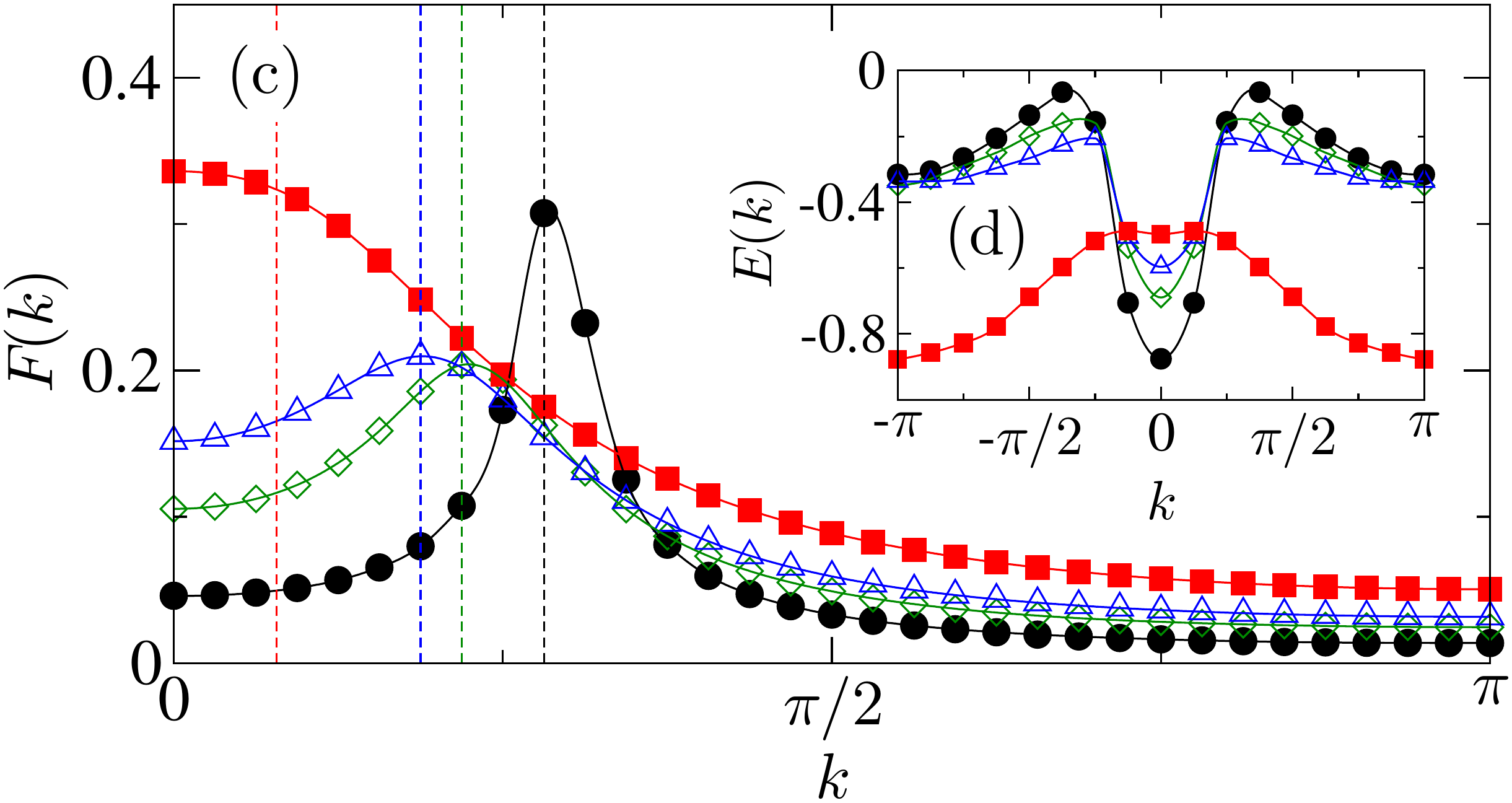}
 \end{center}
 \caption{(Color online) 
 Anomalous spectral function $F(k,\omega)$ in the 1D EFKM with $U=1$ (a)
 and $U=1.9$ (b), where $t_f=-0.1$, $D=1$.  
 Data are obtained by ED using $\eta=0.1$, $L=16$, and PBC.
 Numerical results for $F(k)$ (c) and $E(k)$ (d) are shown 
 for $U=1$ (circles), $1.5$ (diamonds), $1.7$ (triangles), and $1.9$
 (squares). $F(k)$ is determined by the DMRG for $L=64$ (PBC), 
 whereas $E(k)$ is extracted from the lowest peaks of single-particle spectra 
 $A(k,\omega)$ calculated by ED for $L=16$ (PBC).
 Dashed lines in the panel (c) mark the corresponding Fermi momenta
 $k_{\rm F}=\pi N_{c}/L$ in the noninteracting limit.
 }
\label{Fkw-Fk}
\end{figure}
Figures~\ref{Fkw-Fk}(a) and \ref{Fkw-Fk}(b) show the anomalous spectral function
$F(k,\omega)$ in the weak-coupling ($U=1$)
and strong-coupling ($U=1.9$) regimes, respectively, where $D=1$. 
In the former case the EI arises from a semimetallic phase. As a consequence 
most of the spectral weight of the quasiparticle excitations
is located around the Fermi points $k=\pm k_{\rm F}$, again indicating a BCS-type pairing of electrons and holes. 
Obviously, Fermi surface effects play no role for large $U$ where the Hartree shift drives
the system in the semiconducting regime. Here the excitation gap occurs
at $k=0$. 
Note that the gap between the lowest energy peaks in $F(k,\omega)$ is
equal to the binding energy $E_{\rm B}$ given by Eq.~(\ref{E_B-static}).
Figure~\ref{Fkw-Fk}(c) displays  the frequency-integrated quantity $F(k)$.
At $U=1$, $F(k)$ exhibits a sharp peak at the Fermi momentum.
Increasing $U$ the peak weakens and shifts to smaller momenta.
Close to the EI-BI transition point $U=1.9\lesssim U_{{\rm c}_2}=1.92$,
$F(k)$ has a maximum at $k=0$ but is spread out in momentum space, indicating that
the radius of electron-hole pairs becomes small in real space.
Panel (d) gives the quasiparticle dispersion $E(k)$ derived from $A(k,\omega)$.
Driving the BCS-BEC crossover by increasing $U$, the peaks around $k=\pm k_{\rm F}$
disappear as well as the notch around $k=0$. 
Instead a valence band with a flat top around $k=0$ develops, 
just as observed  e.g. in quasi-1D Ta$_2$NiSe$_5$~\cite{WSTMANTKNT09}.

\begin{figure}[tbp]
 \begin{center}
 \includegraphics[clip,width=\columnwidth]{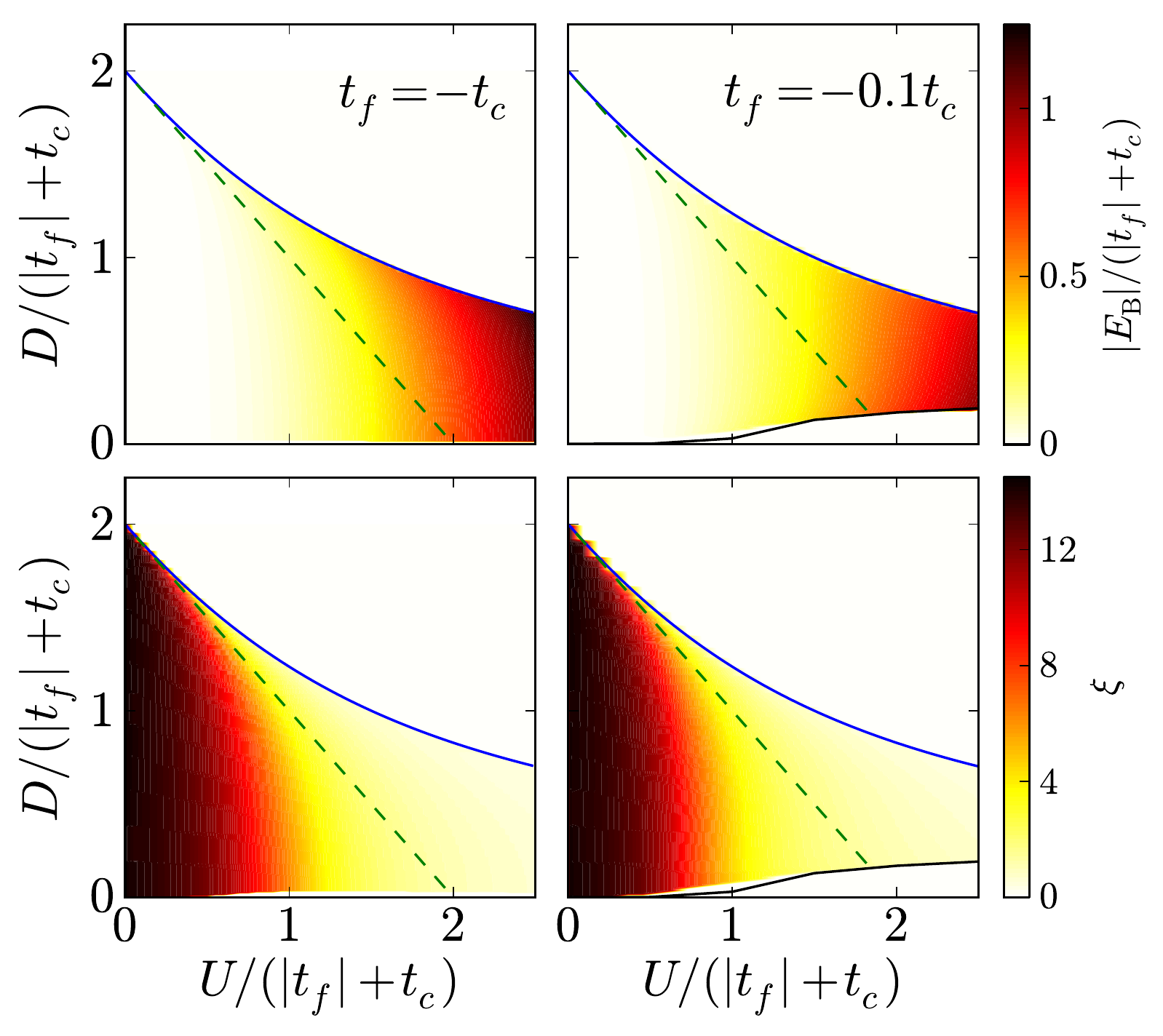}
 \end{center}
 \caption{(Color online) Intensity plots of the binding energy 
 $E_{\rm B}$  (upper panels; $L=128$, OBC)
 and the coherence length $\xi$  (lower panels; $L=64$, PBC)
 in the rescaled $U/(|t_f|+t_c)$--$D/(|t_f|+t_c)$ plane. Data were
 calculated by the DMRG for $N_f>L/2$ 
 (to avoid the AF state in the Hubbard model limit
 $|t_f|=1$, $D=0$). Solid lines denote the SOO-EI and EI-BI transition points
 in the thermodynamic limit (in the lower panels the small uncolored  slot just above the SOO-EI appears because
 $|E_{\rm B}|$ and $\xi$ are obtained here for a fixed finite system
 size). 
 The dashed line [$U_{\rm BI}(D)$] would separate the semimetallic and 
 semiconducting phases if the EI is assumed to be absent.
 }
\label{PD_rescaled}
\end{figure}

Figure~\ref{PD_rescaled} shows the variation of the coherence length 
and the binding energy in the EI phase of the 1D EFKM 
with $|t_f|=1$ (left panels) and $0.1$ (right panels). 
At small $U$ the excitonic state is composed of 
electron-hole pairs having large spatial extension, leading to large values of $\xi$. 
$E_{\rm B}$, on the other hand, is rather small, but increases
exponentially with $U$. This typifies a BCS pairing mechanism. 
At large $U$, the binding increases linearly with $U$. 
Here, tightly bound spatially confined excitons acquire 
quantum coherence (with $\xi\ll 1$) in a Bose-Einstein condensation process.

We finally address the influence of a mass imbalance between $f$- and $c$-band
quasiparticles. The EI phase is absent for $t_f=0$. 
In the mass-symmetric case $|t_f|=t_c$, the 1D Hubbard model results for
$D=0$. Here we cannot distinguish between the AF (with vanishing spin
gap) and EI phases, 
because both phases are critical. 
Therefore, in this limit, we have examined the 1D EFKM for $N_f>L/2$. 
To this end, both the $U$ and $D$ axes in Fig.~\ref{PD_rescaled}
have been rescaled by $(|t_f|+t_c)$, as suggested by the 
EI-BI transition lines~\eqref{Dc2-exact}. Indeed we find that EI phase shrinks as $|t_f|$
decreases. That is, the mass anisotropy gets stronger, which is simply a bandwidth effect, 
however, leading to a stronger Ising anisotropy. This, on their part, enlarges the SOO region, while the EI-BI phase boundary 
basically is unaffected. Importantly, the location of the BCS-BEC crossover, which can
be derived from the intensity plots for $E_{\rm B}$ and $\xi$, 
does not change in this presentation. To expose correlation effects, we included
in Fig.~\ref{PD_rescaled} the semimetallic-to-semiconducting transition
line assuming that the EI phase is absent.
$U_{\rm BI}(D)$ can  be obtained from the band gap $\Delta_{\rm c}$ 
that depends linearly on $U$ for fixed $D$: 
$\Delta_{\rm c}(D)=U+2(|t_f|+t_c)+U_{\rm BI}(D)$ 
[i.e., $U_{\rm BI}(D)$ scales again with $|t_f|+t_c$]. 
Apparently in the BCS-BEC crossover regime a strong renormalization 
of the band structure due to the incipient $f$-$c$ hybridization takes place.

To conclude, adopting the numerically exact density matrix renormalization  
group technique, we examined the one-dimensional (1D) extended Falicov-Kimball model (EFKM) 
and, most notably, proved the excitonic insulator (EI) state shown to be critical.
The complete ground-state phase diagram was derived, and put into 
perspective with the Bethe ansatz results obtained in the strong-coupling limit
for the spin-1/2 XXZ chain. Besides the EI to band insulator transition, 
the boundary between the EI and a phase with staggered orbital ordering
was determined with high accuracy. 
The whole phase diagram of the 1D EFKM could be scaled by
$|t_f|+t_c$; staggered orbital ordering appears 
only for small mass-imbalance ratios $|t_f|/t_c$.
The absence of an order parameter prevents addressing the problem of excitonic condensation in 
1D systems by usual mean-field approaches. That is why we exploited the off-diagonal anomalous 
Green function. The related anomalous spectral function elucidates 
the different nature of the electron-hole pairing and condensation process at weak and strong couplings.  
At fixed level splitting the binding energy between $c$ electrons and $f$ holes is exponentially 
small in the weak-coupling regime. It strongly increases 
as the Coulomb attraction increases. Concomitantly the coherence length of the 
electron-hole pair condensate shortens. This unambiguously demonstrates 
a crossover from BCS-like electron-hole pairing to a Bose-Einstein
condensation of preformed excitons. 
The quasiparticle band dispersion in the BEC regime 
exhibits a rather dispersionless valence band near $k=0$, 
despite the fact that the expectation value $\langle c^{\dagger}f^{}\rangle$ 
is zero because of the 1D setting.
This result further supports the EI scenario for quasi-1D Ta$_2$NiSe$_5$, where
the flat valence-band top was detected by ARPES experiments.

The authors would like to thank Y.~Fuji, F.~G\"ohmann, S.~Nishimoto, 
K.~Seki, T.~Shirakawa, and B.~Zenker for valuable discussions.
S.E. and H.F. acknowledge funding by the DFG through SFB 652 Project B5. 
T.K. was supported by a JSPS Research Fellowship for Young Scientists.
Y.O. acknowledges the Japanese Kakenhi Grant No. 22540363.

\bibliography{ref}
\bibliographystyle{apsrev4-1}

\end{document}